\documentclass[oldversion,letter]{aa}

\usepackage{graphicx,natbib,epsfig}

\begin{document}

\title{Clathrate hydrates as a sink of noble gases in Titan's atmosphere}

\author{C. Thomas, O. Mousis, V. Ballenegger and S. Picaud}

\institute{Institut UTINAM, CNRS-UMR 6213, Universit{\'e} de Franche-Comt{\'e}, France}

\offprints{C. Thomas, \\   e-mail: caroline.thomas@univ-fcomte.fr}

\date{Received / Accepted}

\authorrunning{Caroline Thomas et al.}
\titlerunning{Sink of noble gases in Titan's atmosphere}

\abstract{We use a statistical thermodynamic approach to determine
the composition of clathrate hydrates which may form from a
multiple compound gas whose composition is similar to that of
Titan's atmosphere. Assuming that noble gases are initially
present in this gas phase, we calculate the ratios of xenon,
krypton and argon to species trapped in clathrate hydrates. We
find that these ratios calculated for xenon and krypton are
several orders of magnitude higher than in the coexisting gas at
temperature and pressure conditions close to those of Titan's
present atmosphere at ground level. Furthermore we show that, by
contrast, argon is poorly trapped in these ices. This trapping
mechanism implies that the gas-phase is progressively depleted in
xenon and krypton when the coexisting clathrate hydrates form
whereas the initial abundance of argon remains almost constant.
Our results are thus compatible with the deficiency of Titan's
atmosphere in xenon and krypton measured by the {\it Huygens}
probe during its descent on January 14, 2005. However, in order to
interpret the subsolar abundance of primordial Ar also revealed by
{\it Huygens}, other processes that occurred either during the
formation of Titan or during its evolution must be also invoked.}


\maketitle

\section{Introduction}
\label{intro}

An unexpected feature of the atmosphere of Titan is that no
primordial noble gases other than argon were detected by the Gas
Chromatograph Mass Spectrometer (GCMS) aboard the {\it Huygens}
probe during its descent on January 14, 2005. The detected argon
includes primordial ${}^{36}$Ar (the main isotope) and the
radiogenic isotope ${}^{40}$Ar, which is a decay product of
${}^{40}$K (Niemann et al. 2005). The other primordial noble gases
$^{38}$Ar, Kr and Xe were not detected by the GCMS instrument,
yielding upper limits of 10$^{-8}$ for their mole fractions. In
any case, ${}^{36}$Ar/${}^{14}$N is about six orders of magnitude
lower than the solar value, indicating that the amount of
${}^{36}$Ar is surprisingly poor within Titan's atmosphere
(Niemann et al. 2005).

Several scenarios have been proposed in the literature to
interpret the deficiency of Titan's atmosphere in primordial noble
gases. The first category of scenarios postulates that this
deficiency results from processes that occurred either during
Titan's accretion or during the formation of its planetesimals. In
this way, Alibert \& Mousis (2007) proposed that the building
blocks of Titan initially formed at low temperature in the solar
nebula and were subsequently partially vaporized in Saturn's
subnebula during their migration and accretion, thus leading to
the loss of noble gases and of carbon monoxide\footnote{CO is
several orders of magnitude less abundant than CH$_4$ in Titan's
atmosphere.}. Alternatively, Owen (2006) proposed that Titan's
planetesimals were directly produced at temperatures high enough
in Saturn's subnebula to impede the trapping of the noble gases
during their formation and growth. The second category of
scenarios postulates that Titan initially incorporated primordial
noble gases during its formation but that subsequent processes in
its atmosphere or on its surface prevented them from being
detected during the descent of the {\it Huygens} probe. For
example, Jacovi et al. (2005) suggested that the aerosols observed
in Titan's atmosphere may have cleared its content in noble gases,
assuming they were produced from the aggregation of polymers.

Another interpretation is the trapping of atmospheric compounds in
clathrate hydrates (Osegovic \& Max 2005 -- hereafter OM05).
Indeed, in such ice structures, water molecules form cages which
are stabilized by the trapping of volatiles. Thus, OM05 calculated
clathrate hydrate compositions on the surface of Titan by using
the program CSMHYD developed by Sloan (1998) and showed that such
crystalline ice structures may act as a sink for Xe. However, from
these results and because no other noble gases are considered in
this program, they assumed similar trapping efficiencies for Ar
and Kr. In addition, because the CSMHYD code is not suitable below
160 K for gas mixtures relevant to the atmospheric composition of
Titan, OM05 also extrapolated their results to the surface
temperature of Titan (about 94 K). Moreover, the stability curves
of multiple guest clathrate hydrates calculated from the CSMHYD
code strongly depart from the experimental data at low
temperature.

Here, we reinvestigate OM05's assumptions and calculate more
accurately the trapping of noble gases in clathrate hydrates using
a statistical thermodynamic model based on experimental data and
on the original work of van der Waals \& Platteeuw (1959). In
particular, the relative abundances of Xe, Kr, and Ar are
explicitly calculated in conditions that are valid for the present
temperature and pressure at the surface of Titan. Our results show
that, when clathration occurs at low temperature and pressure, the
trapping of Xe and Kr is very efficient, contrary to that of Ar.
This conclusion is partly consistent with the {\it Huygens} probe
measurements, and supports indirectly the presence of noble
gas-rich clathrate hydrates on the surface of Titan.

\section{Clathrate hydrates composition}
\label{comp}

We follow the method described by Lunine \& Stevenson (1985) to
calculate the abundance of guests incorporated in clathrate
hydrates from a coexisting gas of specified composition. This
method uses classical statistical mechanics to relate the
macroscopic thermodynamic properties of clathrate hydrates to the
molecular structure and interaction energies. It is based on the
original ideas of van der Waals \& Platteeuw (1959) for clathrate
formation, which assume that trapping of guest molecules into
cages corresponds to the three-dimensional generalization of ideal
localized adsorption. In this model, the occupancy fraction of a
guest molecule $K$ for a given type $t$ of cage ($t$~=~small or
large) in a given type of clathrate hydrate structure (I or II)
can be written as

\begin{equation}
\label{occupation}
y_{K,t}=\frac{C_{K,t}P_K}{1+\sum_{J}C_{J,t}P_J} ,
\end{equation}

\noindent where the sum at the denominator includes all the
species which are present in the initial gas phase. $C_{K,t}$ is
the Langmuir constant of species $K$ in the cage of type $t$, and
$P_K$ is the partial pressure of species $K$. This partial
pressure is given by $P_K=x_K\times P$, with $x_K$ the molar
fraction of species $K$ in the initial gas phase, and $P$ the
total pressure.

The Langmuir constants are determined by integrating the molecular potential within the cavity as

\begin{equation}
\label{langmuir}
C_{K,t}=\frac{4\pi}{k_B
T}\int_{0}^{R_c}\exp\Big(-\frac{w(r)}{k_B T}\Big)r^2dr ,
\end{equation}

\noindent where $R_c$ represents the radius of the cavity assumed
to be spherical, and $w(r)$ is the spherically averaged Kihara
potential representing the interactions between the guest
molecules and the H$_2$O molecules forming the surrounding cage.
Following McKoy \& Sinano\u glu (1963), this potential $w(r)$ can
be written for a spherical guest molecule, as

\begin{eqnarray}
\label{pot_Kihara}
w(r)&=&2z\epsilon\frac{\sigma^{12}}{R_c^{11}r}\Big(\delta^{10}(r)+\frac{a}{R_c}\delta^{11}(r)\Big)\nonumber\\
&-&\frac{\sigma^6}{R_c^5
r}\Big(\delta^4(r)+\frac{a}{R_c}\delta^5(r)\Big) ,
\end{eqnarray}

\noindent with

\begin{equation}
\delta^N(r)=\frac{1}{N}\Big[\Big(1-\frac{r}{R_c}-\frac{a}{R_c}\Big)^{-N}-\Big(1+\frac{r}{R_c}-\frac{a}{R_c}\Big)^{-N}\Big].
\end{equation}

\noindent The Kihara parameters $a$, $\sigma$ and $\epsilon$ for
the molecule-water interactions of the species considered here are
given in Table \ref{tab:ParamKihara}. In Eq. (\ref{pot_Kihara}),
$z$ is the coordination number of the cell and $r$ the distance of
the guest molecule from the cavity center. These parameters are
given in Table \ref{tab:ParamCages} and depend on the structure of
the clathrate and on the type of the cage (small or large). Note
that the Kihara parameters are based on fit to experimental data.
However, a systematic study of the influence of their values on
the present results would be required to assess the relevance of
the present conclusions (Papadimitriou et al. 2006).

\begin{table}[h]
\centering
\caption{Parameters of Kihara potential: $a$ is the radius of the impenetrable core, $\epsilon$ is the depth of the potential well, and $\sigma$ is the Lennard-Jones diameter.
}
\begin{tabular}{lcccc }
\hline
\hline
Molecule        & $\sigma$(\AA)  & $a$(\AA) & $\epsilon/k_B$(K) & Ref. \\
\hline
CH$_4$      & 3.1514        & 0.3834        & 154.88        & (a)\\
C$_2$H$_6$  & 3.2422        & 0.5651        & 189.08        & (a)\\
N$_2$           & 3.0224        & 0.3526        & 127.67        & (a)\\
Ar              & 2.829         & 0.226         & 155.30        & (b)\\
Kr              & 3.094         & 0.307         & 212.70        & (b)\\
Xe              & 3.3215        & 0.2357        & 192.95        & (a)\\
\hline
\end{tabular}\\
\parbox{6.5cm}{
\begin{tiny}
(a) Jager (2001)\\
(b) Diaz Pe\~na et al. (1982)
\end{tiny}
}
\label{tab:ParamKihara}
\end{table}

\begin{table}[h]
\centering
\caption{Parameters for the cavities. $b$ represents the number of small ($b_s$) or large ($b_\ell$) cages per unit cell for a given structure of clathrate, $R_c$ is the radius of the cavity (parameters taken from Sparks et al. 1999), and $z$ the coordination number in a cavity.}
\begin{tabular}{lcccc}
\hline
\hline
Clathrate structure & \multicolumn{2}{c}{I} & \multicolumn{2}{c}{II} \\
\hline
Cavity type     & small     & large     & small     & large \\
$b$                     & 2         & 6         & 16        & 8     \\
$R_c$(\AA)      & 3.905     & 4.326     & 3.902     & 4,682 \\
$z$                     & 20        & 24        & 20        & 28    \\
\hline
\end{tabular}
\label{tab:ParamCages}
\end{table}

Finally, the relative abundance $f_K$ of a guest molecule $K$ in a
clathrate hydrate can be calculated with respect to the whole set
of species considered in the system as

\begin{equation}
\label{abondance} f_K=\frac{b_s y_{K,s}+b_\ell y_{K,\ell}}{b_s \sum_J{y_{J,s}}+b_\ell \sum_J{y_{J,\ell}}},
\end{equation}

\noindent where $b_s$ and $b_\ell$ are the number of small and
large cages per unit cell respectively, for the considered
clathrate hydrate structure (I or II). The sums in the denominator
include all species present in the system. Note that, even if it
has been shown that the size of the cages can be reduced by $\sim$
10-15\% at low temperatures (Shpakov et al. 1998; Besludov et al.
2002), we assume here that this size remains fixed. In the same
way, we neglect possible cage distortion due to large guests.
Moreover, the relative abundances calculated this way do not
discriminate between the various isotopes of the same species.

All calculations have been performed at the dissociation pressure
$P=P^{\rm diss}_{\rm mix}$ of the multiple guest clathrate
hydrate, {\it i.e.} temperature and pressure conditions at which
the clathrate hydrate is formed. This dissociation pressure can be
deduced from the dissociation pressure $P^{\rm diss}_K$ of a pure
clathrate hydrate of species $K$, as (Hand et al. 2006)

\begin{equation}
\label{Pdiss} P^{\rm diss}_{\rm
mix}=\Big(\sum_{J}\frac{x_J}{P^{\rm diss}_J}\Big)^{-1}.
\end{equation}

\noindent The dissociation pressures $P^{\rm diss}_J$ derived from
laboratory measurements follow an Arrhenius law (Miller 1961) as

\begin{equation}
\label{Pdiss_pur} \log(P^{\rm diss})=A+\frac{B}{T} ,
\end{equation}

\noindent where $P^{\rm diss}$ is expressed in Pa and $T$ is the
temperature in K. The constants $A$ and $B$ fit to experimental
data (Lunine \& Stevenson 1985; Sloan 1998) in the present study
are given in Table \ref{tab:coeff_pression}. Note that, in the
present approach, the dissociation pressures and the Langmuir
constants are independently calculated.

\begin{table}[h]
\centering \caption{Parameters of the dissociation curves for
various single guest clathrate hydrates. A is dimensionless and B
is in K.}
\begin{tabular}{lcc}
\hline \hline
Molecule        & $A$           & $B$  \\
\hline
CH$_4$      & 9.89          & $-$951.27  \\
C$_2$H$_6$  & 10.65         & $-$1357.42 \\
N$_2$           & 9.86          & $-$728.58  \\
Ar              & 9.34          & $-$648.79  \\
Kr              & 9.03          & $-$793.72  \\
Xe              & 9.55          & $-$1208.03 \\
\hline
\end{tabular}
\label{tab:coeff_pression}
\end{table}

\section{Results and discussion}
\label{res} We determine the relative abundances $f_{\rm Xe}$,
$f_{\rm Kr}$ and $f_{\rm Ar}$ with respect to all guests
incorporated in clathrate hydrate formed from a gas whose
composition is similar to that of Titan's atmosphere. Similarly to
OM05, we assume that N$_2$, CH$_4$, C$_2$H$_6$ and the considered
noble gas (Xe, Kr or Ar) are the main atmospheric compounds of the
initial gas phase (see Table \ref{atmospheres}). Note that when
the calculations are performed for one noble gas (Xe, Kr or Ar),
the two others are excluded from the initial gas composition. The
gas phase abundance of N$_2$ is taken similar to OM05 and that of
CH$_4$ has been updated from the {\it Huygens} measurements
(Niemann et al. 2005). The atmospheric abundance of ethane has
been fixed to 0.1$\%$ in our calculations. For each considered
noble gas, three initial abundances have been selected in order to
investigate the influence of its initial concentration in the gas
phase, on the clathrate hydrate's trapping efficiency. The
abundance of CH$_4$ is then varied in consequence to preserve the
global volatile budget. Moreover, we have calculated the relative
abundances of noble gases for both clathrate hydrates structures I
and II. Because the conclusions of our calculations are similar
for both structures, we focus here on the results obtained for
structure I only. Note that this structure corresponds to that
derived by OM05 from the CSMHYD program.

\begin{table}[h]
\centering \caption{Initial gas phase abundances considered in the
atmosphere of Titan. For each noble gas (Xe, Kr and Ar), three
different compositions of the initial gas phase are investigated.}
\begin{tabular}{lccc}
\hline
\hline
Molecule     & \multicolumn{3}{c}{Molar fractions (\%)}  \\
\hline
Noble gas          & 0.1    & 0.055     & 0.01 \\
CH$_4$      & 4.8       & 4.845         & 4.89 \\
N$_2$           & 95    & 95            & 95  \\
C$_2$H$_6$  & 0.1   & 0.1       & 0.1  \\
\hline
\end{tabular}
\label{atmospheres}
\end{table}

\begin{figure}
\resizebox{\hsize}{!}{\includegraphics[angle=0]{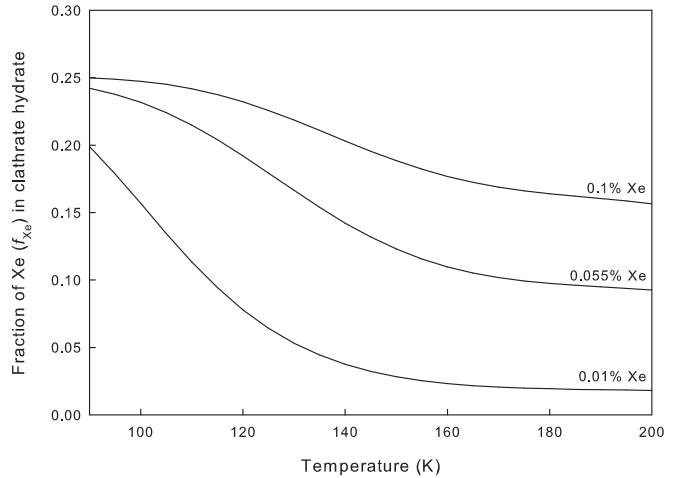}}
\caption{$f_{Xe}$ in a multiple guest clathrate hydrate, as a
function of temperature and of the initial gas phase composition
(see Table \ref{atmospheres}).} \label{Xe}
\end{figure}

\begin{figure}
\resizebox{\hsize}{!}{\includegraphics[angle=0]{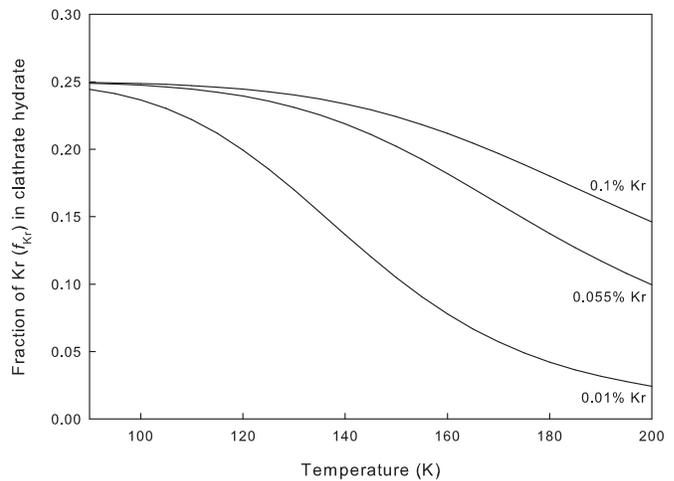}}
\caption{Same as Fig.\ref{Xe} but for Kr.}
\label{Kr}
\end{figure}

\begin{figure}
\resizebox{\hsize}{!}{\includegraphics[angle=0]{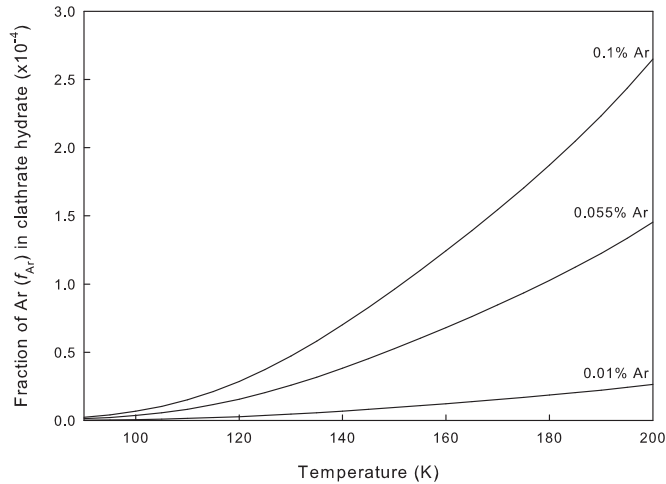}}
\caption{Same as Fig.\ref{Xe} but for Ar.}
\label{Ar}
\end{figure}

Figures 1 to 3 represent the evolution of $f_{{\rm Xe}}$, $f_{{\rm
Kr}}$ and $f_{{\rm Ar}}$ respectively, in a multiple guest
clathrate hydrate as a function of {\it T} along the dissociation
curve $P^{\rm diss}_{\rm mix} = f(T)$, and for three different
noble gas abundances in the initial gas phase (see Table
\ref{atmospheres}). Figure \ref{Xe} shows that, irrespective of
the initial gas phase abundance of Xe, its relative abundance
progressively increases in clathrate hydrate when {\it T}, and
hence $P^{\rm diss}_{\rm mix}$, decrease. This behavior of the Xe
clathration efficiency agrees with the calculations of OM05 (e.g.
their Fig. 5). Similar behavior is seen with Kr, as shown in
Figure \ref{Kr}. By contrast, the relative abundance of Ar in
clathrate hydrate decreases when {\it T} and $P^{\rm diss}_{\rm
mix}$ decrease, as shown in Fig. \ref{Ar}. This figure also shows
that $f_{Ar}$ gradually converges towards zero at temperatures
below $\sim$ 100 K, irrespective of its initial concentration in
the gas phase. This huge difference between the trapping of Ar, Xe
and Kr in clathrate hydrate is mostly due to the higher values of
the Kihara parameters of the two latter molecules (see Table
\ref{tab:ParamKihara}) which are responsible for a stronger
interaction energy with the surrounding water molecules in
clathrate hydrate cages.

\begin{table}[h]
\centering \caption{Abundance ratios of noble gas in clathrate
hydrate to noble gas in the initial gas phase for Xe, Kr and Ar.
These ratios are calculated at {\it T=}171 K and $P^{\rm
diss}_{\rm mix}=$ 1.5 bar.}
\begin{tabular}{lc}
\hline
\hline
 Initial molar fraction         & abundance  \\
 in gas                                     & ratio \\
                        &   \\
\hline
Xe gas                          &  \\
0.001                       & 168 \\
0.00055                         & 184 \\
0.0001                      & 204 \\
\hline
Kr gas                          &  \\
0.001                       & 195 \\
0.00055                         & 285 \\
0.0001                      & 555 \\
\hline
Ar gas                          &  \\
0.001                       & 1.577$\times10^{-1}$ \\
0.00055                     & 1.572$\times10^{-1}$ \\
0.0001                      & 1.566$\times10^{-1}$ \\
\hline
\end{tabular}
\label{tab:resultats}
\end{table}

In addition to the relative abundances $f_K$ discussed above, we
have also calculated the abundance ratio for the three considered
noble gases Xe, Kr and Ar (Table \ref{tab:resultats}). This ratio
is defined as the ratio between the relative abundance $f_K$ of a
given noble gas in the multiple guest clathrate hydrate and its
initial gas phase abundance $x_K$ (see Table~\ref{atmospheres}).
We considered one particular point located on the dissociation
curve of the multiple guest clathrate hydrate under study in our
calculations. This point corresponds to the present atmospheric
pressure at the ground level of Titan ($P^{\rm diss}_{\rm
mix}$~=~1.5 bar) and to a temperature value of 171 K, as indicated
by the clathrate hydrate dissociation curve. Table
\ref{tab:resultats} shows that, depending on the initial gas phase
conditions and for this particular point, the relative abundances
of Xe and Kr trapped in clathrate hydrates are between 168 and 555
times those in the initial gas phase. Considering the current
amount of atmospheric CH$_4$ calculated by Lunine \& Tittemore
(1993) ($\sim$ $3 \times 10^{20}$ g), we have calculated that a
0.1\% gas phase abundance of a given noble gas translates into the
presence of $\sim$ $3.9 \times 10^{17}$ moles in Titan's
atmosphere. Assuming full efficiency for the sink mechanism
described here, the amount of water needed for the trapping of Xe
or Kr in clathrate hydrates at the surface of the satellite is
$\sim$ $4 \times 10^{19}$ g. This roughly corresponds to a single
guest (Xe or Kr) clathrate hydrate layer $\sim$ 50 cm thick at the
surface of Titan.

Such an efficient trapping may be high enough to significantly
decrease the atmospheric concentrations of Xe and Kr, provided
that clathrate hydrates are abundant enough on the surface of
Titan. On the contrary, with an abundance ratio much lower than 1,
Ar is poorly trapped in clathrate hydrate and the Ar atmospheric
abundance consequently remains almost constant.

\section{Conclusions}
\label{con}

The trapping mechanism we investigated can explain the deficiency
in primordial Xe and Kr observed by {\it Huygens} in Titan's
atmosphere. Indeed, our results show that these noble gases could
have been progressively absorbed in clathrate hydrates located at
the surface of Titan during its thermal history. By contrast, our
results show that the trapping of Ar is poor in clathrate
hydrates. As a consequence, such a mechanism alone cannot
interpret the subsolar abundance of primordial Ar which is also
observed in Titan's atmosphere. The different scenarios invoked in
the Introduction may thus have worked together with the trapping
in clathrate hydrates to explain the deficiency in primordial Xe,
Kr and Ar. Note that boulders observed in images from {\it
Huygens} have been invoked as evidence for the existence of
clathrate hydrates on the surface of Titan (OM05). However,
further {\it in situ} measurements are required to investigate
whether noble gases-rich clathrate hydrates really exist on the
surface of Titan.

\begin{acknowledgements}
We acknowledge E. Dendy Sloan and Nicolas Iro for fruitful
discussions and information on their work. We also acknowledge
Gabriel Tobie for useful comments on this manuscript.
\end{acknowledgements}

{}

\clearpage

\end{document}